\documentclass[twocolumn]{aastex701}
\renewcommand{\textbf}[1]{#1}

\hypersetup{
    colorlinks=true,
    linkcolor=blue,
    filecolor=cyan,      
    urlcolor=magenta,
    citecolor=blue,
}
\usepackage{adjustbox}
\usepackage{siunitx}
\usepackage{makecell}
\usepackage{booktabs}
\usepackage{natbib}
\usepackage{hyperref}
\usepackage{amsmath}
\usepackage{array}
\usepackage{graphicx}
\usepackage{xcolor}
\usepackage{txfonts}
\usepackage{float}
\usepackage{subfigure}
\usepackage{multirow}
\usepackage{rotating}
\usepackage{balance}
\usepackage[mathlines]{lineno}

\chardef\us=`\_

\submitjournal{ApJ}

\begin{document}

\title{Linking Magnetic Field Diagnostics with 3D CME Speeds in Solar Active Regions}

\author[0000-0003-4772-2492]{Harshita Gandhi}
\affiliation{Department of Physics, Aberystwyth University, Aberystwyth, Wales, UK}
\email[show]{hag43@aber.ac.uk}

\author[0000-0002-6547-5838]{Huw Morgan}
\affiliation{Department of Physics, Aberystwyth University, Aberystwyth, Wales, UK}
\email[show]{hmorgan@aber.ac.uk}




\begin{abstract}
Understanding how active-region properties influence coronal mass ejection (CME) dynamics is essential for constraining eruption models and improving space-weather prediction. Magnetic diagnostics derived above polarity inversion lines (PILs), including the critical height ($h_{\rm crit}$) of torus instability onset, the overlying field strength ($B_{\rm t}$), and ribbon flux ($R_{\rm f}$), provide physically motivated measures of eruption onset. The two main aims of this work are to (i) show that $h_{\rm crit}$ and $B_{\rm t}$ can equally well predict CME speeds when evaluated over the region of interest (ROI) not directly above the PIL, and (ii) assess the value of $h_{\rm crit}$, $B_{\rm t}$ and $R_{\rm f}$ in predicting CME speed. Photospheric magnetograms are modeled with potential-field extrapolations to obtain decay index profiles. Critical heights above PILs correlate strongly with 3D CME speed ($r = 0.71$). Using ROIs of $\approx$ 1.8, 3.7, and 7.3 Mm), centered on the PIL, weighted $h_{\rm crit}$ from the 7.3x7.3 ROI provides the strongest correlation ($r = 0.73$), while mean $B_{\rm t}$ at 150 Mm is weaker ($r = 0.33$). Combining both offers little improvement ($r = 0.74$), confirming $h_{\rm crit}$ as the dominant predictor. CME speed correlates moderately with $B_{\rm t} \times R_{\rm f}$ ($r = 0.44$), and highest when combined with $h_{\rm crit}$ ($r = 0.76$). Thus, in potential field models, ROI-based critical heights are as predictive as those above the PIL, indicating that the broader active-region field structure is equally valid as a diagnostic. When all parameters are considered together, $h_{\rm crit}$ alone consistently shows the highest predictive power for CME speed.
\end{abstract}


\section{Introduction}
\label{sec:intro}
Eruptive activity on the Sun is driven by the release of magnetic free energy stored in non-potential coronal fields within active regions \citep{low1996solar, forbes2000review}. Energy is accumulated through flux emergence, shear flows, or flux cancellation at the photosphere, processes that form highly sheared arcades or magnetic flux ropes \citep[e.g.,][]{green2009flux, green2011photospheric, green2013observations}. The stability of these structures depends on the balance between the upward Lorentz force of the stressed core field and the restraining tension of the overlying coronal field. \textbf{A loss of equilibrium occurs when the upward force exceeds this confinement, leading to the eruption of the magnetic structure.  Such eruptions often, though not always, expel plasma and magnetic flux into the heliosphere as coronal mass ejections (CMEs).}

\textbf{Depending on the mechanism, this imbalance may arise from ideal processes, such as the torus instability, or from resistive processes, such as magnetic reconnection in a current sheet beneath the flux rope that weakens the overlying field} \citep{forbes1995photospheric, forbes2006cme,chen2011coronal}. In the torus instability scenario, loss of equilbrium occurs when the upward hoop force of a toroidal flux rope exceeds the downward tension of the overlying poloidal (‘strapping’) field \citep{kliem2006torus, deng2017roles}. \textbf{The condition is quantified by the magnetic decay index, which describes how rapidly the strength of the overlying magnetic field decreases with height:
\begin{equation}
n(h) = -\frac{d \ln B_{\text{ext,p}}}{d \ln R},
\label{eq:decayindex}
\end{equation}
where $B_{\text{ext,p}}$ is the external poloidal field and $R$ (or equivalently, height $h$) is the radial distance above the photosphere.} A larger $n$ indicates a faster decline of the overlying field and therefore weaker confinement of the flux rope. Instability occurs when $n$ exceeds a critical threshold $n_c$ (typically 1–2, canonical $\sim$1.5; \textbf{\citep{bateman1978mhd, fan2007onset,aulanier2010formation,xu2012relationship, aggarwal2018prediction, zuccarello2015critical, rees20202d}}), defining the critical height, $h_{\rm crit}$. Above this height, steeper $n(h)$ profiles imply faster decay of the restraining transverse field, $B_{\rm t}$, leading to stronger acceleration, suggesting a link between the slope of $n(h)$ above $h_{\rm crit}$ and CME speed \citep{kliem2006torus, kliem2025magnetic}. \textbf{In resistive scenarios, reconnection can add poloidal flux to the rope, observed through flare ribbons. The ribbon flux (hereafter denoted $R_{\rm f}$) serves as a proxy for the total reconnected magnetic flux which contributes to the acceleration of the CME during its impulsive phase.} Thus, CME dynamics is expected to depend on the decay index profile $n(h)$, the critical height $h_{\rm crit}$, the strength of the overlying transverse field $B_{\rm t}$, and the reconnected flux $R_{\rm f}$ \citep{james2022evolution, gandhi2025linking, wang2017critical, cheng2020initiation, aulanier2009formation}.

CME dynamics is controlled by the evolving coronal magnetic field, which sets the stability of flux ropes and governs their onset and acceleration. This makes the pre-eruption magnetic configuration a key determinant of what triggers an eruption and how rapidly it develops. Understanding these connections is central to solar physics and space-weather prediction, as CME kinematics in the low corona, particularly during the acceleration phase, drive shock formation, particle acceleration, and geomagnetic storm intensity \citep{forbes2006cme, schmieder2015flare, chen2017physics, majumdar2022variation}. Most CMEs complete their acceleration below $\sim$1~$R_{\odot}$ \citep{macqueen1983kinematics, zhang2006statistical, vrvsnak2008processes}, making early estimates of CME speed from source-region diagnostics both valuable and challenging for operational forecasting.

Observational studies link CME speed to several magnetic parameters. Decay-index profiles below $\sim$40 Mm show positive correlation with CME speed up to $\sim$1000 km s$^{-1}$ before saturating at higher values \citep{xu2012relationship}. Ribbon flux has also been identified as a key driver, showing stronger correlations than decay index \citep{deng2017roles}. Steep $n(h)$ slopes are associated with fast halo CMEs, whereas broad dips in $n(h)$ correspond to slower events, consistent with simulations \citep{kliem2024decay}. Other studies emphasize the role of critical height: \citet{gandhi2025linking} reported a 71\% correlation between CME speed and $h_{\rm crit}$ when Polarity inversion lines (PILs) were identified automatically, compared to weaker correlations using plane-of-sky speeds or manual PIL selection. \citet{james2022evolution} showed that eruptions are more likely to occur when critical heights are rising, particularly during periods of flux emergence. \citet{kim2017relation} reported that 3-D CME speeds correlate more strongly with helicity-injection rate and total unsigned flux than with 2-D speeds, with the correlations becoming weaker in active regions where the helicity sign reversed. Regions with more free energy, larger PILs, and strong transverse fields tend to produce faster CMEs \citep{cui2018statistical}. The flux-accretion model of \citet{welsch2018flux} proposes that reconnection adds poloidal flux to the erupting rope, increasing acceleration. Meanwhile, findings presented by \citet{sherolli2023cme} suggest that, after allowing for ribbon flux, regions with stronger coronal field strength at the eruption site may further influence CME speed. Prominence eruptions may require lower decay-index thresholds (0.8–1.3) and accelerate more strongly when associated with flares \citep{vasantharaju2019finding}. Geometric factors can alter instability thresholds. For short, low-lying flux ropes, the critical decay index can fall well below $n_c \approx 1.5$ \citep{filippov2021critical}. Nonlinear force-free field (NLFFF) modeling shows that critical height distinguishes eruptive from confined flares more effectively than helicity ratios \citep{gupta2024stability}. Case studies confirm that torus instability and reconnection can act together: filaments and overlying loops may accelerate simultaneously \citep{schrijver2008observations}, and tether-cutting reconnection during confined flares can form a flux rope that later erupts once $n(h)$ exceeds a threshold \citep{kliem2021nonequilibrium}.

Despite these advances, many studies have relied on plane-of-sky CME speeds, which underestimate true CME velocities and weaken correlations with source-region parameters. Analyses often restrict field extrapolations to manually selected or partial PILs, with critical-height and decay-index measurements frequently evaluated at a single height. These constraints can miss the influence of surrounding fields, contributing to the modest ($\lesssim0.6$) correlations often reported and the lack of a consistent framework linking magnetic properties to CME speed.

This study builds on \citet{gandhi2025linking} (hereafter Paper I), analyzing 28 events. Potential-field extrapolations are done and extrapolated field is used to derive decay-index profiles and subsequent critical heights through two approaches. First, PIL segments are automatically detected, and a decay-index profile is constructed from the extrapolated field along the longest high-gradient segment to extract a single critical height. Second, fixed regions of interest (ROIs: 5×5, 10×10, 20×20 pixels, corresponding to physical sizes of $\approx$ 1.8, 3.7, and 7.3 Mm on a side and $\approx$ 2.5$^{\prime\prime}$, 5.0$^{\prime\prime}$ and 10.0$^{\prime\prime}$) and full region centered on the PIL yield critical-height metrics (mean, median, max and \textbf{$B_{t}$}-weighted) and mean transverse field strength at multiple heights (40, 70, 100 and 150Mm). Correlations with 3D CME speeds test whether magnetic diagnostics can predict CME speeds, and whether additional factors like field strength and ribbon flux contribute alongside critical height. This provides a direct test of the flux-accretion hypothesis and evaluates whether ROI-based methods reproduce PIL-tracking results.
\begin{figure*}
    \centering
    \includegraphics[width=0.75\linewidth]{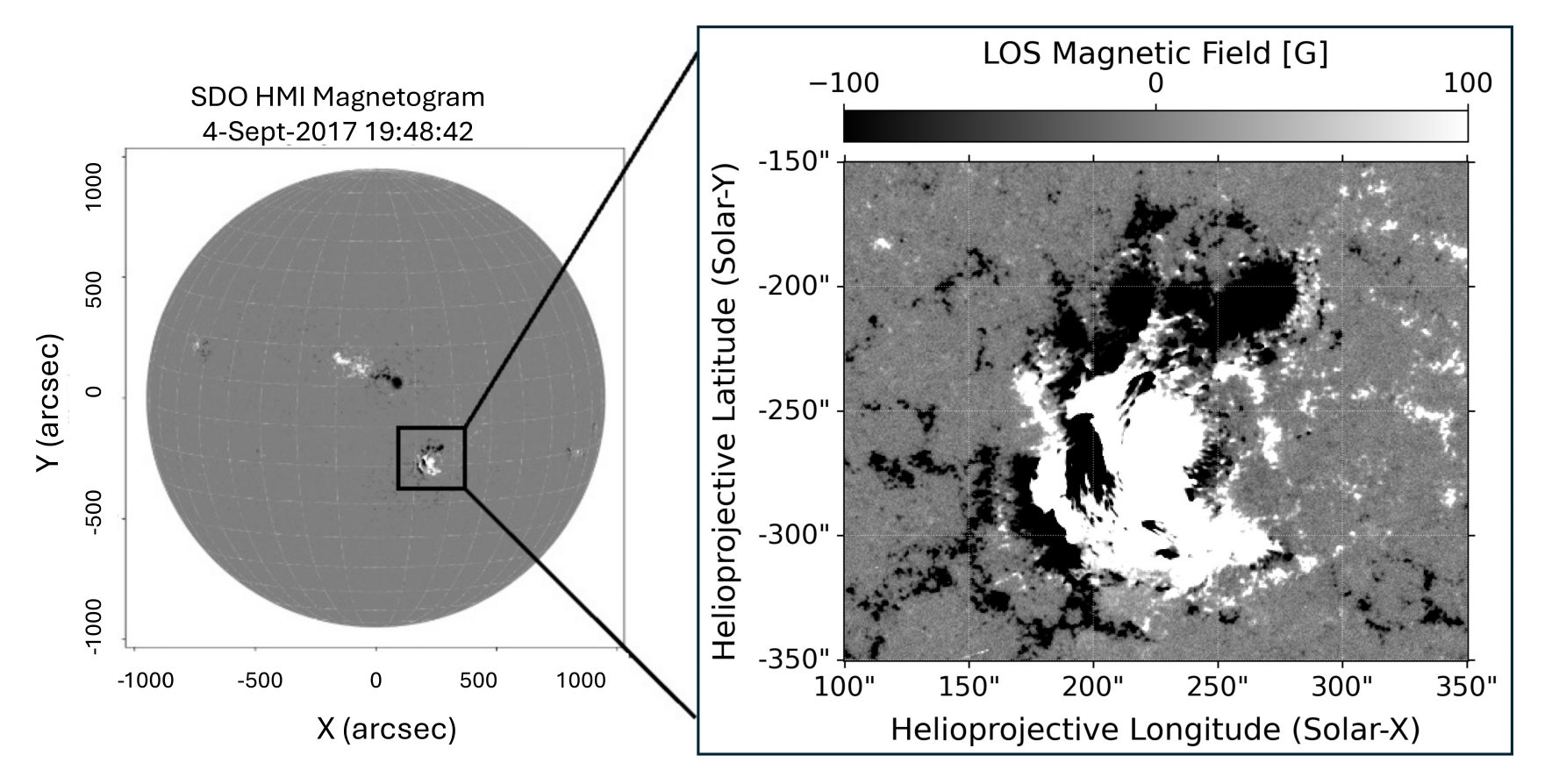}
    \caption{Full disk line-of-sight magnetogram observed by HMI onboard SDO on 2017 September 4 at 19:48:42 UT with zoomed in cutout of the active region used for the application of the extrapolation method.}
    \label{fig:fig1}
\end{figure*}
The findings verify and extend Paper I, while aiming to clarify the relative roles of these parameters and provide a more consistent framework for estimating CME speeds from pre-eruption magnetograms.

The paper is organized as follows: Section~\ref{sec:data&method} outlines data and methodology; Section~\ref{sec:Results} presents the results; Section~\ref{sec:Dis} discusses implications and limitations; Section~\ref{sec:conc} summarizes the main conclusions.

\section{Data and Method}
    \label{sec:data&method}
\subsection{Data Selection and Preparation}
    \label{sec:DS&P}
We analyze 28 halo CMEs from active regions, including 24 events from the list defined in Paper~I (see \cite{gandhi2025linking} for details of the selection criteria) and four additional high-speed ($>$1500 km s$^{-1}$) halo events not analysed in Paper I. For the present analysis, we restrict to events from 2010 onwards to ensure uniform coverage with Helioseismic and Magnetic Imager (HMI) data, thereby maintaining consistency in the data source. Beyond this selection, all data preparation and analysis are performed independently. 

\textbf{For each event, we use line-of-sight (LOS) magnetograms from the \texttt{hmi.M\_45s} series, which provides full-disk observations at 45~s cadence. The \texttt{hmi.M\_45s} series is well suited to study rapidly evolving active-region fields and form the basis for the PIL detection and potential field extrapolations. For the PIL-based analysis, we construct hourly time series magnetograms to perform the extrapolations. This approach ensures consistent temporal sampling across events while capturing the photospheric magnetic evolution in the immediate pre- and post-eruption period. In contrast, for the ROI-based analysis, we use a single 45s frame per event, taken at the time closest to the eruption onset.}

In the following, we describe the processing steps applied to HMI data for both PIL-only and ROI-based approaches:

\begin{enumerate}
    \item \textbf{Pre-processing and alignment:}HMI magnetograms contain placeholder values for missing pixels, which are set to NaNs. Each magnetogram is rotated by $180^{\circ}$ to correct the north–south alignment, and the WCS header is updated accordingly.
    
    \item \textbf{Active-region cutout:}The magnetogram is cropped to the bounding box of the source active region, defined in helioprojective Cartesian coordinates. 
    
    \item \textbf{LOS-to-radial conversion:} Assuming the photospheric field is predominantly radial, the heliocentric angle $\theta$ per pixel from the projected distance $r$ to disk center is given by
    \[
    \theta = \arcsin\!\left(\frac{r}{R_\odot}\right), \qquad r = \sqrt{(x-x_c)^2 + (y-y_c)^2},
    \]
    where $R_\odot$ is the apparent solar radius and $(x_c,y_c)$ is the disk center. The radial component is obtained from
    \[
    B_r = \frac{B_{\mathrm{LOS}}}{\cos\theta},
    \]
    \textbf{To avoid numerical instability in the LOS-to-radial conversion, we exclude only those pixels where $\cos\theta < 10^{-6}$, corresponding to a narrow ring ($\sim 0.001^{\circ}$) at the extreme limb. No additional limb threshold is applied, as all active regions in this study are restricted to within $\pm45^{\circ}$ heliographic longitude, where projection effects remain small and the assumption of a predominantly radial photospheric field remains valid.}
    
    \item \textbf{Spatial rebinning} For computational efficiency, the magnetogram cutouts are rebinned by a factor of 3, so that each output pixel corresponds to the mean of $3\times 3$ input pixels, reducing both $x$ and $y$ dimensions by a factor of three while preserving the large-scale field structure. The WCS metadata are updated to maintain the correct angular scaling.
\end{enumerate}

The resulting $B_r$ cutouts as shown in Figure~\ref{fig:fig1}, together with coordinate grids from the updated WCS, provide the input for PIL detection and potential-field extrapolation, from which decay-index profiles and critical heights are derived.
\begin{figure}[h]
    \centering
    \includegraphics[width=\linewidth]{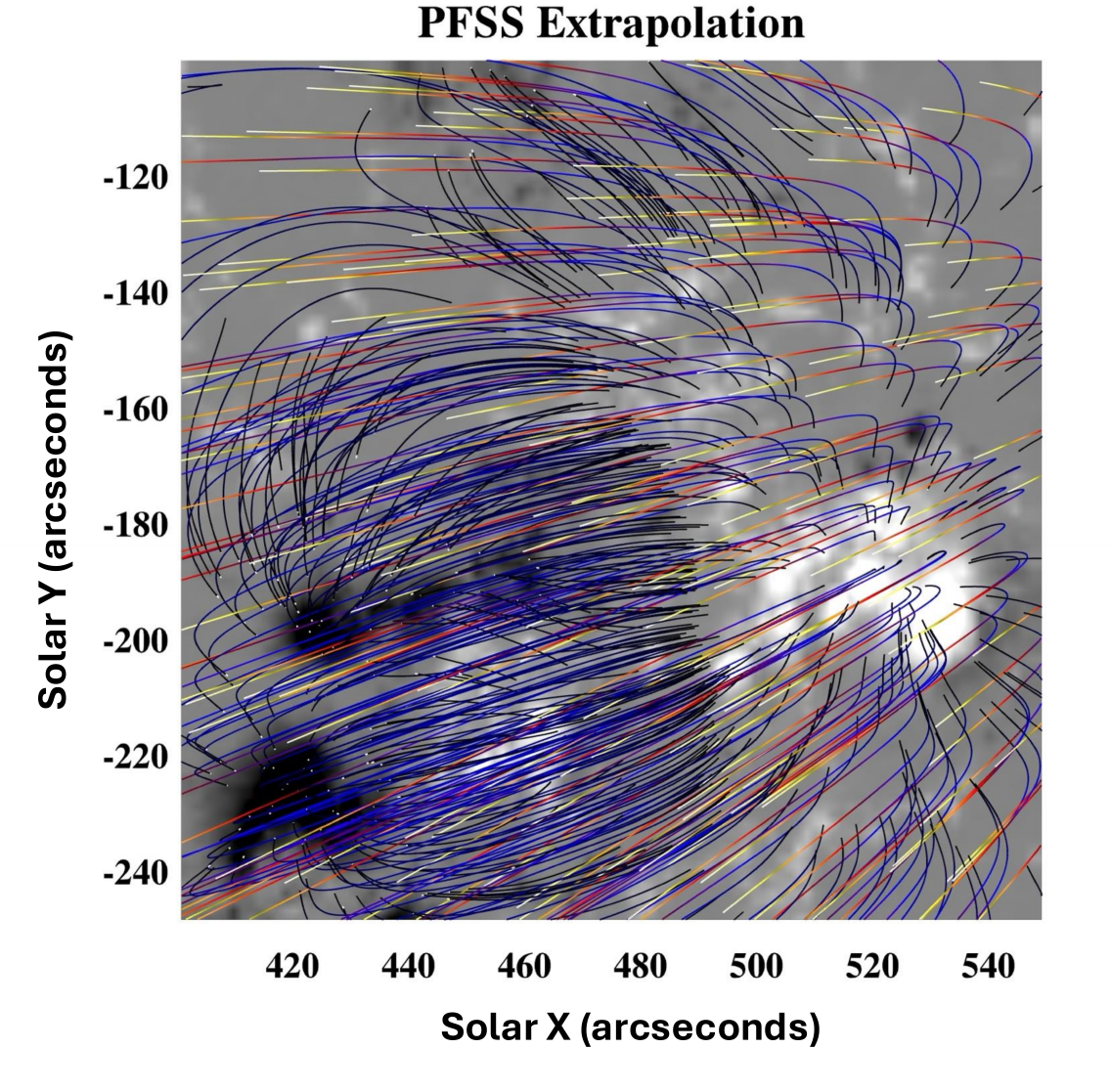}
    \caption{Potential-field extrapolation for one of the studied active regions. The background shows the radial photospheric field from HMI, and coloured curves are traced field lines.}
    \label{fig:pfss_example}
\end{figure}
\subsection{Magnetic field extrapolation using Green’s-function}
    \label{sec:MFE}
The method used in this work is based on a Green’s‐function approach for potential‐field extrapolation from photospheric boundary data. The coronal magnetic field is assumed to be current-free, such that
\[
\nabla \times \mathbf{B} = 0, \qquad \mathbf{B} = -\nabla \Phi ,
\]
with the scalar potential \(\Phi\) satisfying Laplace’s equation,
\[
\nabla^{2} \Phi = 0 .
\]
The lower boundary condition is given by the radial photospheric field \(B_r(x,y,z=0)\) from HMI observations (Section~\ref{sec:data&method}). The computational domain is a rectangular Cartesian volume enclosing the active region with an additional lateral margin to reduce edge effects. The domain extends from the photosphere (\(z=0\)) to a maximum height \(z_{\max}\) (here \(1\,R_\odot\) above the surface), sampled on a uniform vertical grid of order a few hundred layers.

The scalar potential $\Phi$ at each grid point $\mathbf{r}=(x,y,z)$ is evaluated using the Green’s-function solution
\[
\Phi(\mathbf{r}) \;=\; \sum_i \frac{B_{r,i}\,\Delta S_i}{d_i},
\]
where $B_{r,i}$ is the radial field in pixel $i$ (Gauss), $\Delta S_i$ is the pixel area (m$^2$), and $d_i$ is the distance (m) between $\mathbf{r}$ and the centre of pixel $i$. Explicitly,
\[
d_i \;=\; \sqrt{(x-x_i)^2 + (y-y_i)^2 + (z+\delta z)^2}.
\]
\begin{figure*}
    \centering  \includegraphics[width=\linewidth]{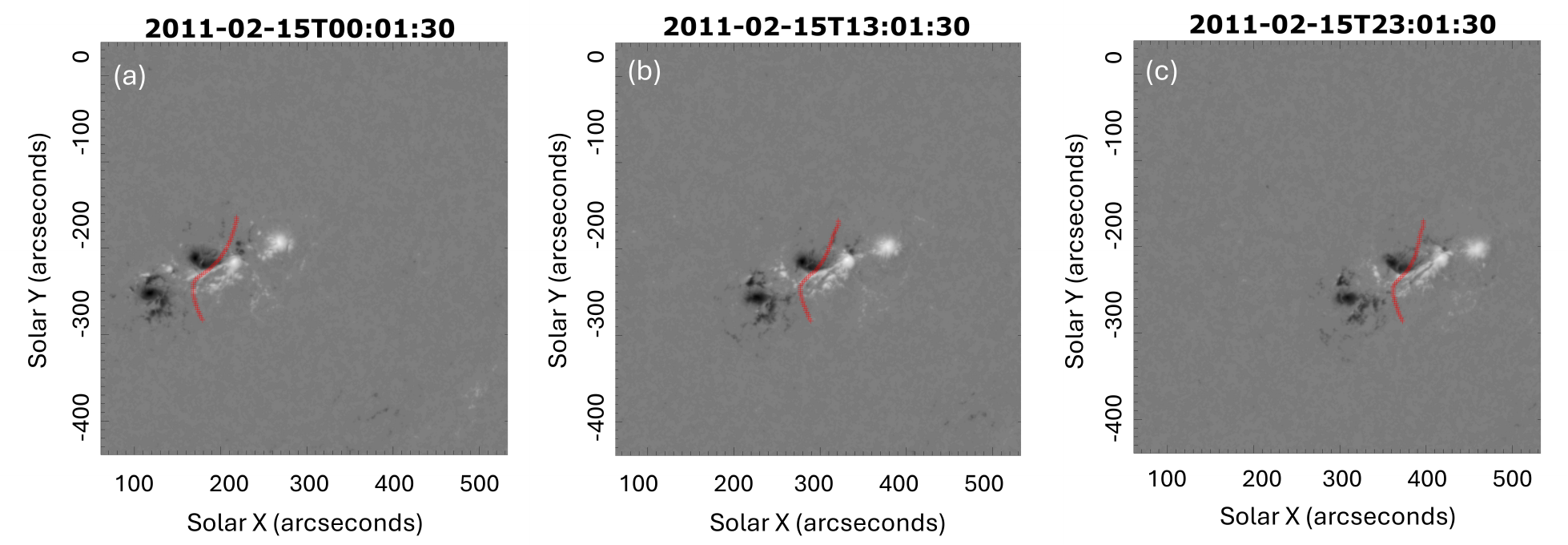}
    \caption{Figure shows HMI $B_r$ cutouts with the longest detected PIL overplotted in red. The temporal sequence shows how the PIL is tracked across magnetograms over time.}
    \label{fig:pil}
\end{figure*}
\begin{figure*}
    \centering \includegraphics[width=0.95\textwidth]{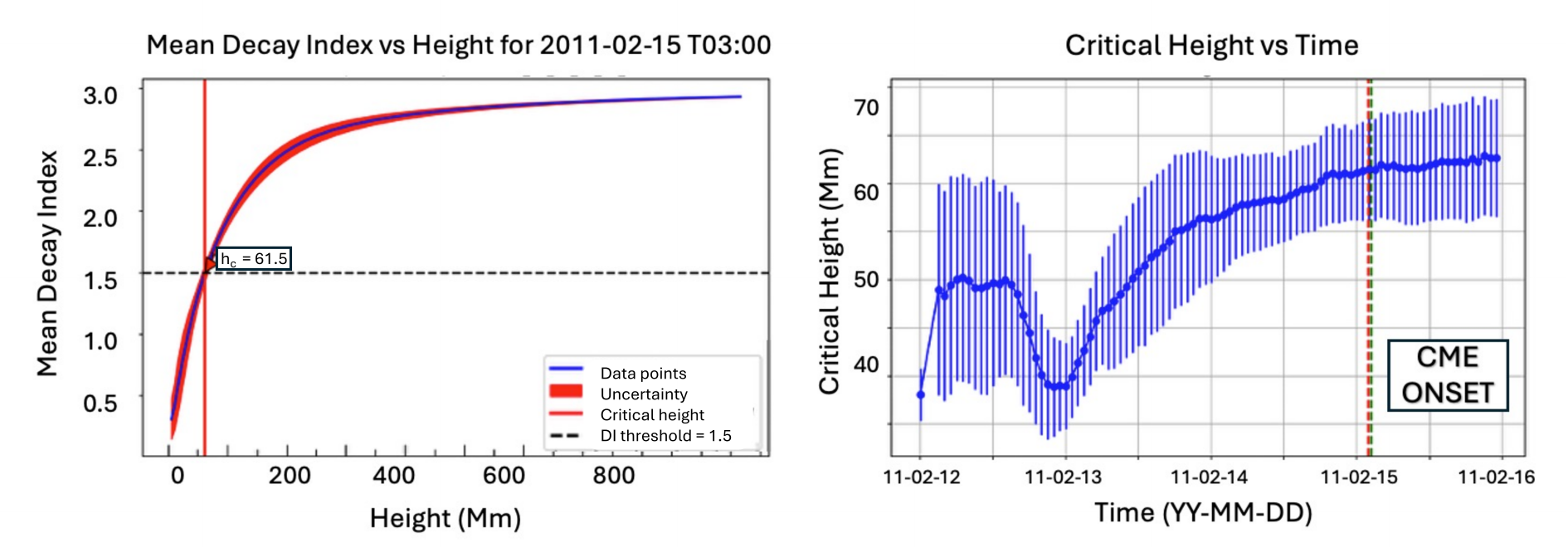}
    \caption{(a) Mean decay index profile along the PIL as a function of height.(b) Temporal evolution of the critical height, with red and blue dashed lines marking CME onset time and the first appearance time in SOHO/LASCO-C2}
    \label{fig:DI&ch}
\end{figure*}
This represents the Euclidean separation from each photospheric source pixel, with a small vertical offset $\delta z$ applied to regularise the $1/d$ kernel at the lower boundary and mimic a finite source-cell height. The pixel area is obtained from the plate scale $\alpha$ (arcsec\,$R_\odot^{-1}$) as
\[
\Delta S_i = \left( \Delta x \,\frac{R_\odot}{\alpha} \right)
             \left( \Delta y \,\frac{R_\odot}{\alpha} \right).
\]
The scalar potential is smoothed in $x$, $y$, and $z$ with Gaussian kernels to suppress grid-scale noise. The magnetic field components are then computed as
\[
B_x = -\frac{\partial \Phi}{\partial x}, \quad
B_y = -\frac{\partial \Phi}{\partial y}, \quad
B_z = -\frac{\partial \Phi}{\partial z},
\]
and the total field strength is
\[
|\mathbf{B}| = \sqrt{B_x^2 + B_y^2 + B_z^2}.
\]
Field lines are traced using the normalised vector field $\mathbf{B}/|\mathbf{B}|$, and lateral margins are removed from the final volume. An example extrapolation with traced field lines over a photospheric magnetogram is shown in Figure~\ref{fig:pfss_example}.

\subsubsection{PIL-only field extrapolation and critical height estimation}
In the first application, the extrapolation is restricted to points located above a detected PIL. The PIL is identified from the photospheric $B_r$ map using a two-step procedure. First, the magnetogram is Gaussian-smoothed to suppress small-scale noise, and pixels where the smoothed field is within $\pm$10 G of zero are flagged as inversion-line locations. Second, these pixels are retained only where the Gaussian-smoothed absolute field exceeds 80 G, so that inversion lines are detected only between regions of strong opposite polarity \citep[cf.][]{schrijver2007R}. The resulting binary mask is thinned and segmented into distinct regions, and the longest continuous PIL is selected (Figure~\ref{fig:pil}).

\textbf{The Green’s-function solution for the potential field (Section~\ref{sec:MFE}) is evaluated only along the vertical columns above the selected PIL using the full photospheric $B_r$ distribution as sources.} This ensures that the derived diagnostics reflect the local strapping field at the eruption site while retaining the influence of the surrounding flux. The transverse field above each PIL pixel is calculated as
\[
B_t(h) = \sqrt{B_x^2(h) + B_y^2(h)}.
\]
A decay-index profile $n(h)$ (Eq.~\ref{eq:decayindex}) is then derived for each vertical column, and the values are averaged along the PIL to obtain a mean profile as a function of height (Figure~\ref{fig:DI&ch}a). The critical height $h_c$ is defined as the first height where $n(h) > n_{\mathrm{crit}}$ (with $n_{\mathrm{crit}}=1.5$). Repeating this procedure for successive magnetogram frames allows the temporal evolution of $h_c$ to be tracked relative to CME onset (Figure~\ref{fig:DI&ch}b).
\begin{figure*}
    \centering \includegraphics[width=0.85\textwidth]{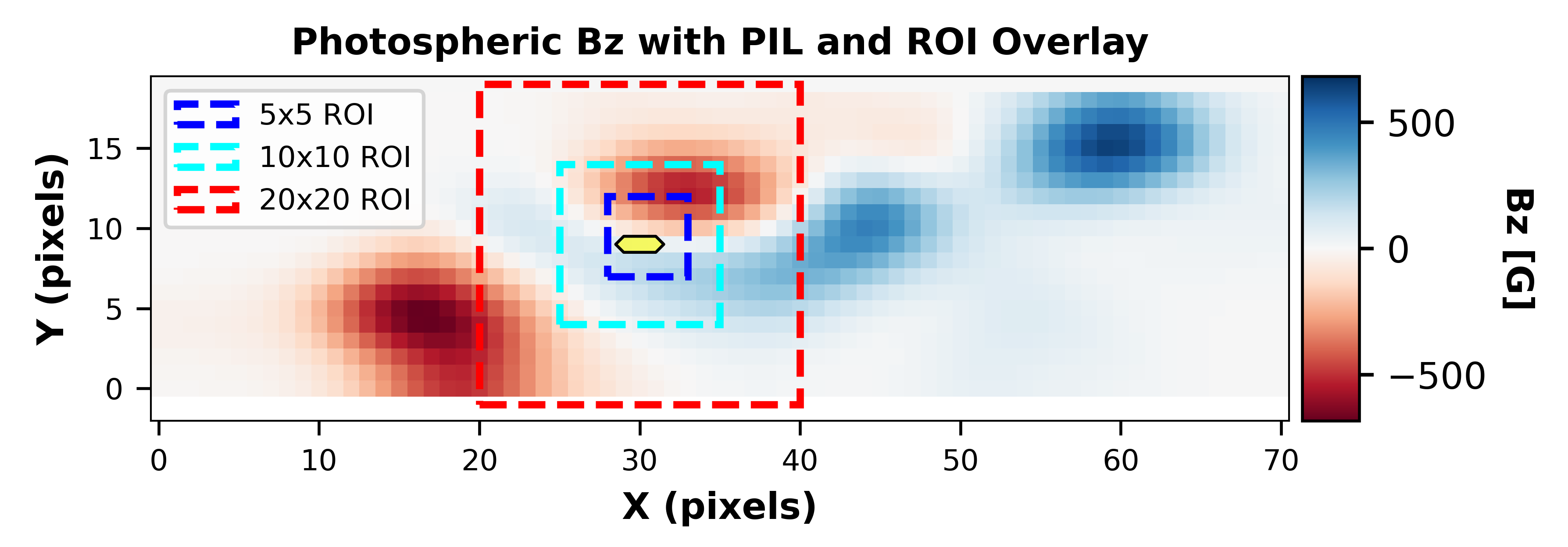}
    \caption{Photospheric $B_z$ map with polarity inversion line (PIL, yellow) and overlaid regions of interest (ROIs): 5×5 pixels (blue), 10×10 pixels (green), and 20×20 pixels (red).}
    \label{fig:fig5}
\end{figure*}
\begin{figure*}
    \centering   \includegraphics[width=0.6\textwidth]{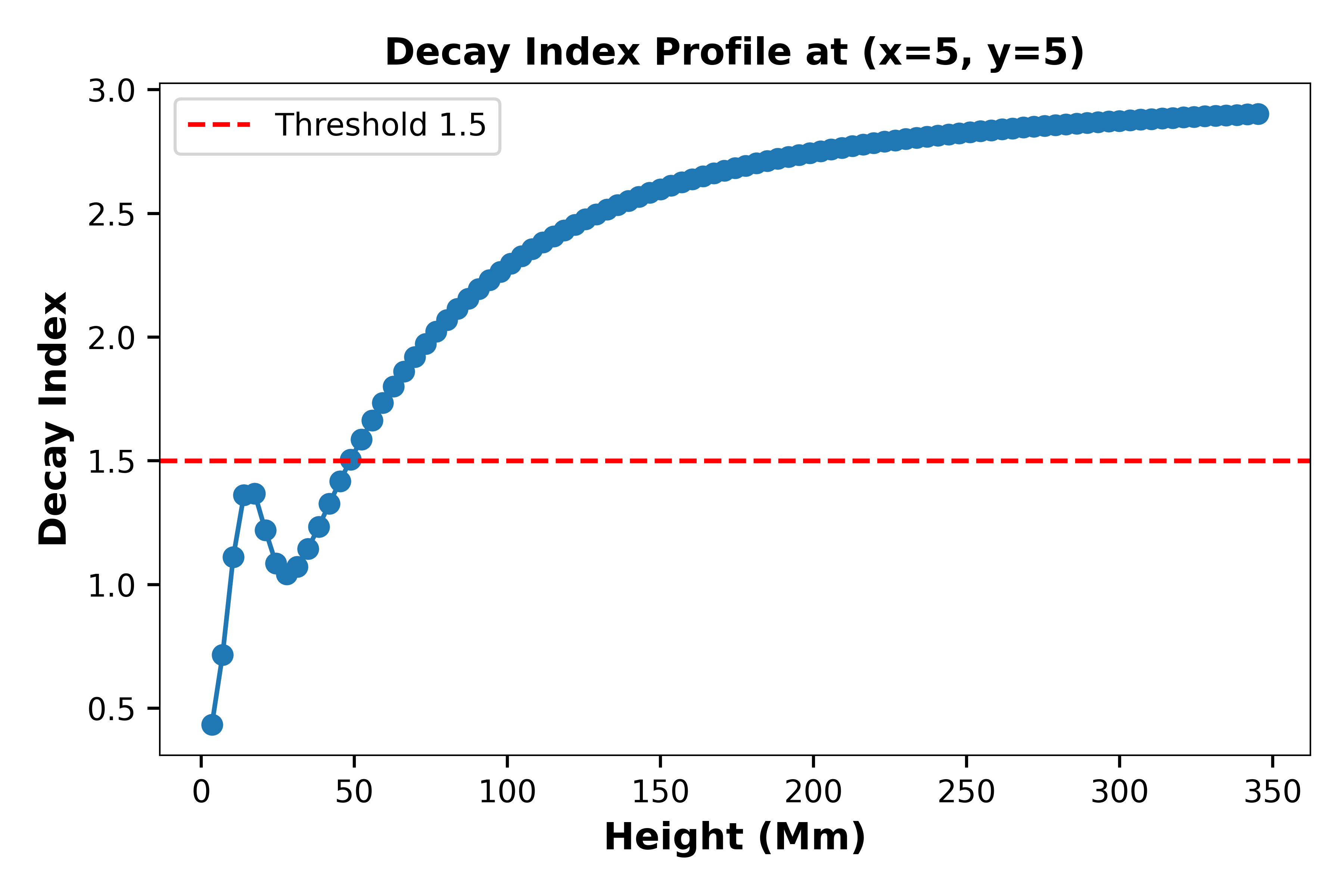}
    \caption{Decay index profile at the central PIL pixel, with \( n = 1.5 \) marked by the red dashed line with \( h_c = 51 \,\mathrm{Mm} \).}
    \label{fig:di}
\end{figure*}
\subsubsection{PIL-centred full region extrapolation and critical height estimation}
\label{sec:PILROI}

In the second application, potential-field extrapolation is performed once for the entire cropped active-region magnetogram just before CME onset, using the photospheric $B_r$ as the lower boundary. The extrapolation provides the full 3D magnetic field components, $B_x(x,y,h)$, $B_y(x,y,h)$, and $B_z(x,y,h)$, throughout the coronal volume. From these, the transverse field strength ($B_t(x,y,h)$) is obtained at each pixel and height as
\[
B_t(x,y,h) = \sqrt{B_x^2(x,y,h) + B_y^2(x,y,h)}.
\]
The role of the PIL is different here than in the PIL-only method. Instead of restricting the extrapolation to PIL pixels, the PIL is used only to locate the center of the regions of interest (ROIs). The PIL is identified using the same automated procedure as in Section~\ref{sec:PIL_only}, and the centroid of the longest continuous segment defines the ROI centre (shown in Figure~\ref{fig:fig5} with overlaid ROIs.)

Square ROIs of $5\times5$, $10\times10$, and $20\times20$ pixels (corresponding to physical sizes of $\approx$ 1.8, 3.7, and 7.3 Mm on a side) are defined around this centroid, together with the full active-region cutout as a fourth case. The full region refers to the entire cropped magnetogram enclosing the active region, without restriction to a localised box. At each pixel within the ROI or full region, the decay index is calculated as in Eq.~\ref{eq:decayindex} from $B_t(x,y,h)$, evaluated over heights up to an upper limit of 350~Mm. The decay index profile at the center of the 10x10 ROI is shown in Figure~\ref{fig:di}.

\begin{figure*}
    \centering    \includegraphics[width=0.85\textwidth]{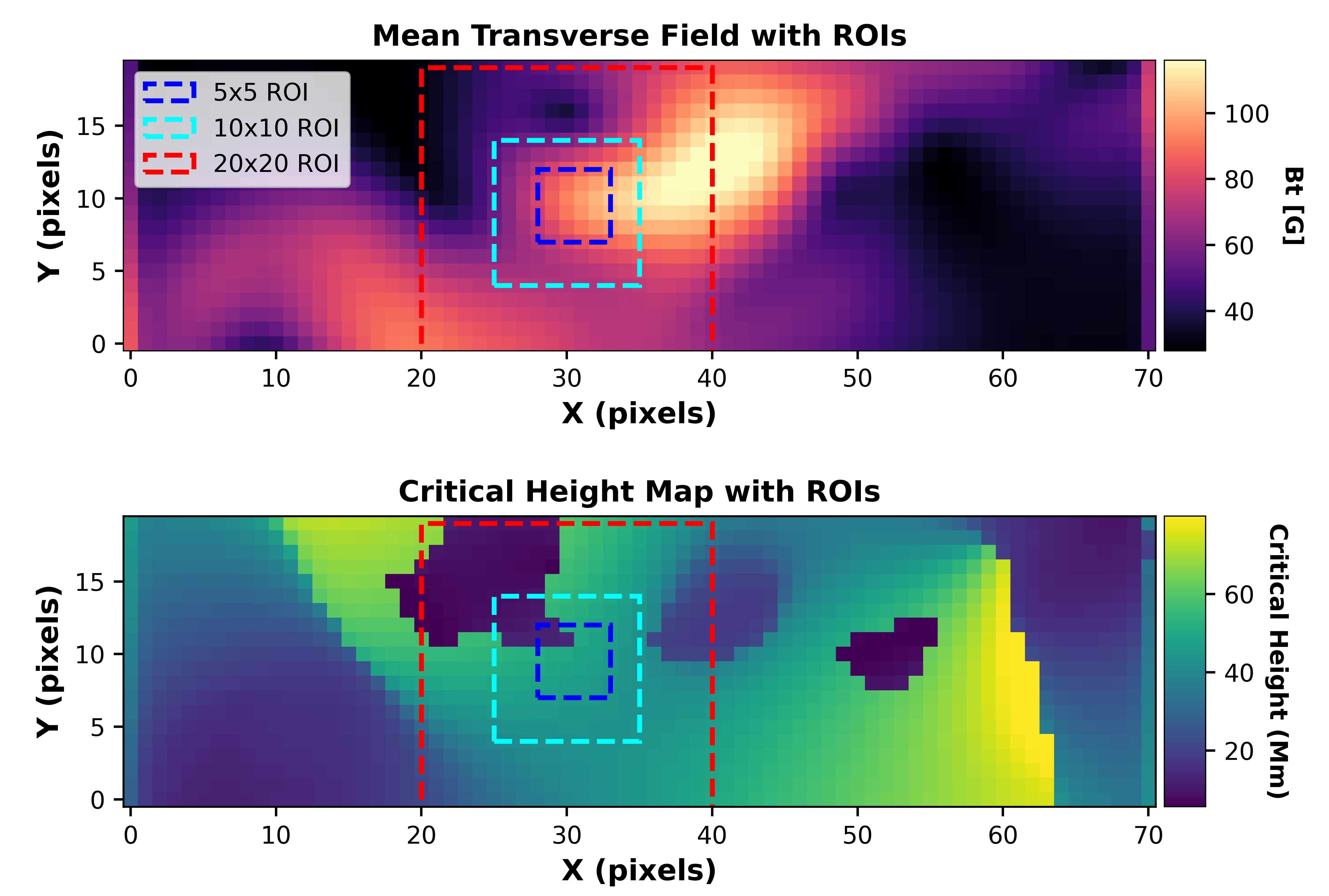}
    \caption{top: average transverse magnetic field strength at each pixel across all heights. bottom: height at which decay index reaches 1.5 — i.e., the critical height for instability at each pixel}
    \label{fig:ch-bt}
\end{figure*}

The critical height $h_c(x,y)$ is defined where $n(h)$ first exceeds 1.5 at each pixel. From the resulting $h_c(x,y)$ maps (shown in Figure~\ref{fig:ch-bt}b), four critical height metrics are extracted for each ROI and full region: maximum, mean, median and a transverse-field-weighted mean critical height as: 

\begin{align*}
\text{Weighted } h_c &: \;\; h_c^{\mathrm{w}} = \frac{\sum h_c(x,y)\,\overline{B_t}(x,y)}{\sum \overline{B_t}(x,y)},
\end{align*}
where $\overline{B_t}(x,y)$ is the height-averaged transverse field at each pixel computed as,
\[
\overline{B_t}(h) = \frac{1}{N_{\mathrm{ROI}}}\sum_{(x,y)\in \mathrm{ROI}} B_t(x,y,h).
\]
is shown in Figure~\ref{fig:ch-bt}a.
\begin{table*}
\small
\hspace*{-0.60in}
\resizebox{550pt}{!}{%
\begin{tabular}{rrrrrrrrrrrrr}
\toprule
NOAA & CME & \makecell{Speed\\(km/s)} & \makecell{PIL-only\\$h_c$} & \makecell{Max\\$h_c$} & \makecell{Median\\$h_c$} & \makecell{Mean\\$h_c$} & \makecell{Weighted\\$h_c$} & \makecell{$\overline{B_t}$\\(40 Mm)} & \makecell{$\overline{B_t}$\\(70 Mm)} & \makecell{$\overline{B_t}$\\(100 Mm)} & \makecell{$\overline{B_t}$\\(150 Mm)} & Ribbon Flux (Mx) \\
\midrule
11092 & 1 & 1260.0 & 92 & 107.9 & 91.8 & 91.8 & 90.7 & 73.9 & 39.4 & 23.1 & 11.3 & 2.22e+20  \\
11158 & 1 & 592.0 & 45.4 & 61.0 & 43.0 & 32.0 & 29.8 & 147.0 & 55.8 & 25.0 & 9.3 & 5.12e+21 \\
11158 & 2 & 480.0 & 39.6 &57.3 & 40.3 & 35.2 & 31.2 & 179.2 & 67.0 & 30.2 & 11.3 & 1.37e+21 \\
11158 & 3 & 494.0 & 35.7 &58.9 & 42.7 & 37.4 & 36.4 & 199.0 & 76.3 & 34.9 & 13.2 & 2.72e+21 \\
11158 & 4 & 627.0 & 37.2 &61.7 & 46.1 & 43.0 & 40.4 & 187.6 & 72.7 & 33.2 & 12.6 & 1.55e+21 \\
11158 & 5 & 876.0 & 41.3&59.4 & 47.6 & 45.0 & 44.8 & 196.3 & 75.7 & 34.7 & 13.2 & 1.16e+22 \\
11166 & 1 & 965.0 & 46.1 &91.9 & 57.6 & 52.3 & 51.2 & 203.6 & 74.0 & 34.8 & 13.9 & 5.18e+21 \\
11261 & 1 & 600.0 & 29.5 &84.5 & 37.8 & 35.7 & 34.6 & 116.5 & 45.6 & 24.7 & 11.2 & 7.07e+21 \\
11261 & 2 & 925.0 & 47.2 &98.1 & 38.6 & 56.5 & 55.8 & 92.2 & 47.1 & 25.4 & 11.0 & 7.61e+21 \\
11261 & 3 & 1200.0 & 51.5 &75.5 & 48.9 & 55.5 & 58.9 & 133.8 & 58.9 & 29.5 & 12.2 & 8.26e+21 \\
11283 & 1 & 855.0 & 39.2 &86.0 & 33.4 & 39.9 & 37.7 & 113.6 & 47.3 & 26.7 & 12.6 & 3.26e+21 \\
11283 & 2 & 782.0 & 32.3 &56.6 & 34.0 & 31.0 & 30.9 & 123.2 & 44.2 & 21.8 & 9.2 & 5.92e+21 \\
11283 & 3 & 810.0 & 33.3 &86.1 & 34.8 & 29.8 & 29.4 & 136.3 & 50.6 & 24.3 & 9.8 & 7.98e+21 \\
11402 & 1 & 735.0 & 31.4 &47.5 & 38.6 & 35.9 & 34.9 & 237.6 & 72.7 & 30.3 & 10.9 & 1.70e+21 \\
11504 & 1 & 755.0 & 38.5 &109.6 & 51.4 & 48.1 & 47.8 & 134.9 & 45.6 & 25.3 & 12.4 & 2.21e+21 \\
11504 & 2 & 999.0 & 54.6 &77.5 & 65.7 & 65.9 & 65.8 & 249.6 & 115.9 & 60.7 & 26.2 & 3.88e+21 \\
11515 & 1 & 939.0 &  44.5 &55.4 & 45.3 & 42.6 & 40.9 & 321.0 & 104.9 & 44.3 & 15.9 & 2.60e+21 \\
11515 & 2 & 740.0 & 34.6 &45.0 & 38.7 & 37.3 & 36.6 & 130.3 & 34.5 & 16.6 & 9.6 & 8.62e+21 \\
11515 & 3 & 990.0 & 42.3 &53.0 & 43.4 & 42.3 & 41.5 & 440.9 & 162.9 & 75.3 & 29.7 & 1.07e+21 \\
11543 & 1 & 650.0 &37.3 &61.8 & 38.9 & 40.3 & 38.4 & 241.8 & 84.6 & 37.5 & 14.1 & 1.04e+21 \\
11618 & 1 & 680.0 & 38.2 &57.4 & 47.7 & 43.6 & 42.2 & 191.0 & 62.6 & 27.0 & 10.0 & 5.34e+21 \\
11875 & 1 & 957.0 & 49.5 &75.5 & 66.0 & 63.6 & 61.8 & 278.7 & 134.4 & 68.9 & 28.7 & 5.34e+21 \\
11974 & 1 & 726.0 & 41.3 &81.3 & 58.5 & 60.0 & 59.7 & 158.7 & 67.9 & 35.9 & 16.3 & 6.69e+21 \\
12036 & 1 & 910.0 & 45.2 &51.5 & 39.5 & 40.1 & 43.3 & 189.5 & 63.8 & 26.9 & 9.6 & 1.66e+21 \\
\textcolor{red}{12673} & 1 & 1850.0 & 53.3 &48.8 & 27.4 & 28.0 & 26.1 & 199.4 & 55.4 & 21.9 & 7.6 & --- \\
\textcolor{red}{12887} & 1 & 1510.0 & 48 &55.7 & 20.6 & 21.7 & 21.0 & 82.8 & 19.3 & 6.8 & 2.1 & --- \\
\textcolor{red}{12891} & 1 & 1530.0 & 59.4 &56.6 & 36.2 & 87.3 & 36.5 & 158.3 & 53.2 & 22.7 & 8.2 & --- \\
\textcolor{red}{11429} & 1 & 2036.0 & 51.5 &60.1 & 31.5 & 31.9 & 30.4 & 277.0 & 87.3 & 36.0 & 12.7 & --- \\
\bottomrule
\end{tabular}}
\caption{CME speeds derived from the GCS reconstructions presented in Paper~I, shown alongside PIL-only and ROI-based magnetic diagnostics for the $20\times20$ region. Critical heights ($h_c$) are in Mm and mean transverse field strengths ($\overline{B_t}$) are in G.}
\label{tab:roi20x20}
\end{table*}
Alongside these maps of $h_c(x,y)$, we also compute mean transverse-field strength $\overline{B_t}(h)$ at fixed coronal heights, $h=40, 70, 100,$ and $150$~Mm. Unlike the PIL-only method, which yields a single representative $n(h)$ profile, the ROI method uses a single pre-eruption magnetogram and characterises the broader magnetic environment surrounding the PIL. The resulting statistical summaries for each ROI and full region are compared with CME speeds in Section~\ref{sec:Results}.

\section{Results}
    \label{sec:Results}
    \vspace{1em}
\subsection{3D CME speeds vs critical heights from PIL-only extrapolation}
\label{sec:PIL_only}

\begin{figure}[h]
    \centering
    \includegraphics[width=0.9\linewidth]{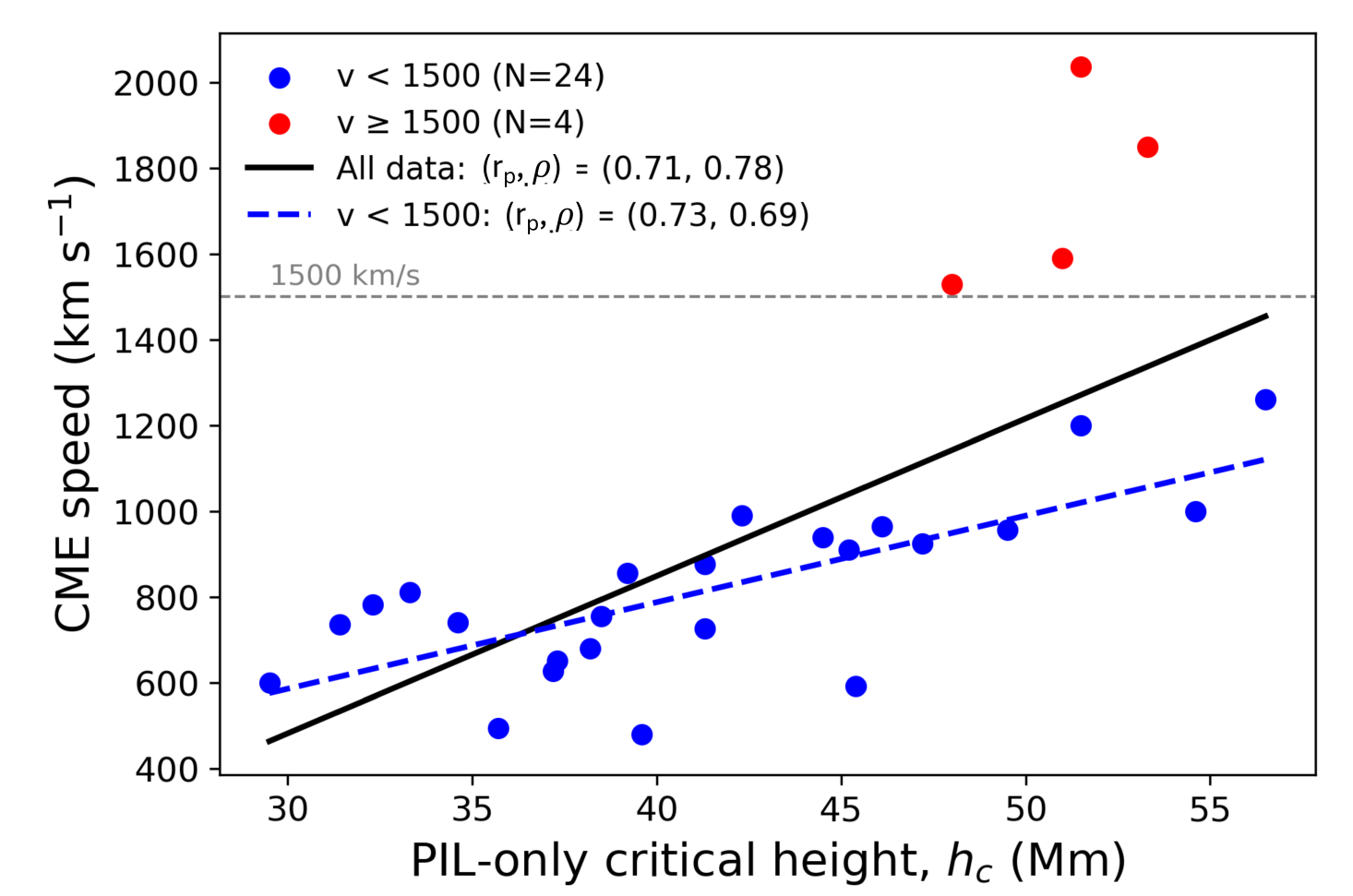}
    \caption{Plot of CME speed as a function of PIL-only critical height. Blue points mark CMEs with speeds below 1500~km~s$^{-1}$, while red points mark the four faster events. The black line shows the linear best-fit using all events, and the dashed blue line shows the fit restricted to the 24 slower events. Pearson ($r$)and Spearman ($\rho$) correlation coefficients for both regressions are shown.}
    \label{fig:fig8}
\end{figure}

Critical heights closest to eruption onset are derived from decay-index profiles evaluated above the longest automatically detected PIL segment in the magnetograms as described in Section~\ref{sec:PIL_only} and shown in Figure~\ref{fig:DI&ch} a and b. Figure~\ref{fig:fig8} shows CME speed as a function of critical height (\( h_c \)) for 28 events. Each point corresponds to a single eruption, with blue symbols marking the 24 CMEs with speeds below $1500$kms$^{-1}$ and red symbols indicating the four faster events. The black solid line shows the best-fit linear regression using all events, giving Pearson and Spearman correlations of $r=0.71$ and $\rho=0.78$, respectively. The dashed blue line shows the regression restricted to the slower subset ($v<1500$kms$^{-1}$), where the correlations are $r=0.73$ and $\rho=0.69$. The figure illustrates a clear trend in which lower critical heights are generally associated with slower CMEs, while events with $h_c > 50$~Mm tend to reach speeds above $1000$kms$^{-1}$.

Since Paper~I used 37 events with speeds below 1500~km~s$^{-1}$ and reported Pearson and Spearman correlations of $0.71 \pm 0.08$ using a similar automated PIL-tracking method, but with a different extrapolation approach based on the Fourier method of \citet{alissandrakis1981field}, we therefore limit the comparison to the 24 events below 1500, matching the speed range. These results confirm that $h_c$ remains a strong single-parameter predictor of CME speed when consistently measured across events. The Pearson correlation obtained in this case ($r=0.73$) is consistent with the value reported in Paper~I, confirming that $h_c$ remains a strong single-parameter predictor of CME speed when consistently measured across events.

\subsection{3D CME speeds vs Critical Heights vs Transverse Field from PIL-centered Full Region Extrapolation} 
For each ROI size and for the full region, we quantify how critical height metrics, transverse field strength at different heights, and ribbon flux relate to CME speed, both individually and in combination. First, we calculate correlation coefficients between CME speed and each critical height metric. Second, we compute the same correlations between CME speed and $\overline{B_t}$ at each of the four heights. Third, we apply a multiple regression of the form
\begin{equation}
    v_{\mathrm{CME}} = a + b\,h_c + c\,\overline{B_t}
    \label{eq:eq2}
\end{equation}
to evaluate the combined effect of critical height and transverse field strength. When all 28 events are included, the four very fast CMEs shown as red points in Figure~\ref{fig:fig10} (v $>$ 1500) act as strong outliers and reduce the correlations to near zero. To avoid this bias, the following results are restricted to the 24 events with v $<$ 1500. These are directly comparable to the PIL-only analysis presented in Section~\ref{sec:PIL_only}. 

\begin{figure*}   \includegraphics[width=\textwidth]{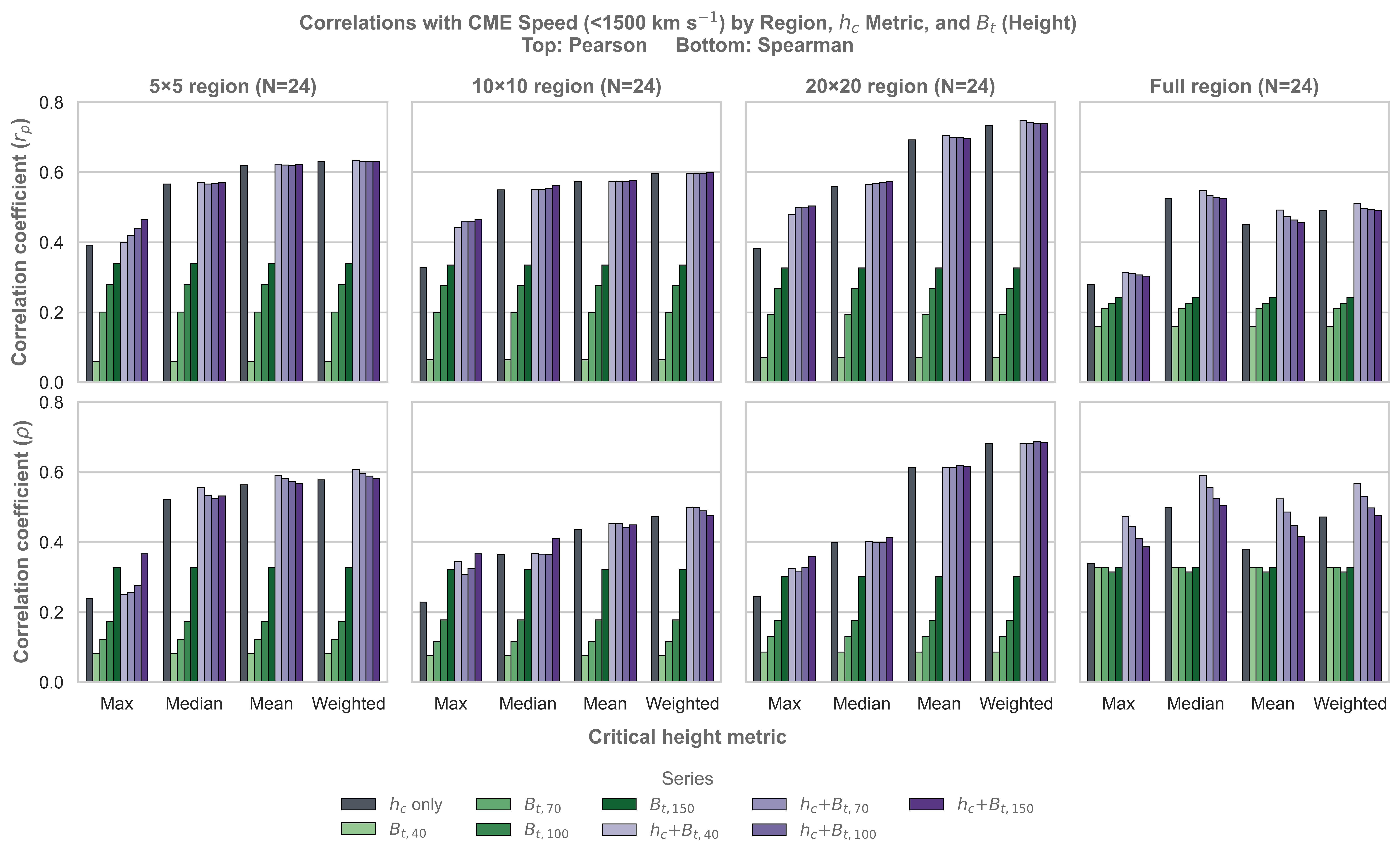}
    \caption{Pearson (top row) and Spearman (bottom row) correlation coefficients between CME speed ($<1500$ km/s) and different predictors. Results are shown for four regions (5×5, 10×10, 20×20, and Full) across four critical height metrics (Max, Median, Mean, Weighted). Bars represent correlations for $h_c$ only, $\overline{B_t}$ at different heights (40, 70, 100, 150 Mm), and combined $h_c$ + $\overline{B_t}$ at respective heights.}
    \label{fig:fig9}
\end{figure*}

\begin{figure}[h]
    \centering   \includegraphics[width=\linewidth]{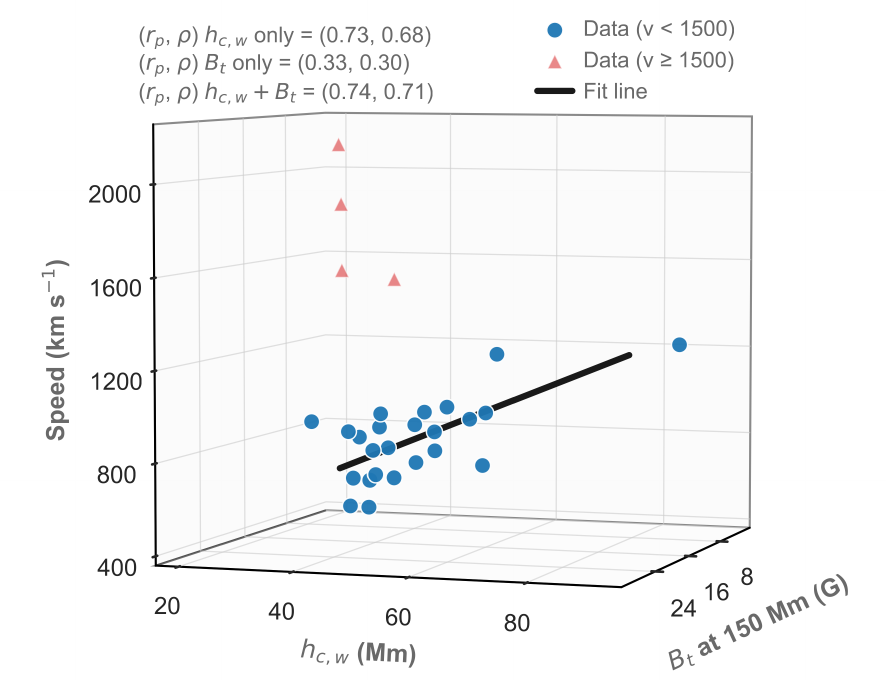}
    \caption{Three-dimensional plot of CME speed versus weighted critical height ($h_c$) and mean transverse field ($\overline{B_t}$) at 150~Mm for the $20\times20$ ROI. Blue points show 24 events with $v<1500$~km~s$^{-1}$, while red points mark the four faster CMEs. The fitted line, based on the slower events, illustrates the combined dependence on $h_c$ and $\overline{B_t}$, following $v = 333.2 + 9.8\,h_c + 2.3\,\overline{B_t}$. Reported correlations correspond to $h_c$ alone, $\overline{B_t}$ alone, and the combined regression.}
    \label{fig:fig10}
\end{figure}

Figure~\ref{fig:fig9} summarises Pearson ($r_p$, top panels) and Spearman ($\rho$, bottom panels) correlation coefficients between CME speed and different predictors, shown separately for four ROI sizes (5$\times$5, 10$\times$10, 20$\times$20 pixels) and the full active-region extent. Each panel compares four critical-height metrics (Max, Median, Mean, Weighted) on the x-axis, with bars representing their correlations with CME speed. \textbf{The dark grey bars correspond to $h_c$ alone, while the green shades denote correlations using the mean transverse field $\overline{B_t}$ at four fixed coronal heights (40, 70, 100, 150 Mm). Purple shades indicate the combined regression results for $h_c$+$B_t$ at these respective heights.} These $B_t$ values are mean fields at the given heights, and are not separate “metrics” in the same sense as the critical heights.

In all cases, correlations with $h_c$ exceed those with $\overline{B_t}$. The monotonic behaviour captured by Spearman’s $\rho$ is consistent with Pearson’s $r_p$, confirming that the observed trends are robust and not dominated by outliers. Correlations with $\overline{B_t}$ alone remain weak across all heights, and adding $\overline{B_t}$ to $h_c$ in a combined regression (purple bars) improves the fit only marginally. The full-region case shows the weakest performance overall, indicating that averaging over the entire active region dilutes the magnetic field contribution from the eruption site.
Among the ROI sizes, the 20x20 pixels gives the strongest correlations, with weighted critical heights performing best and the mean second best. ~Table~\ref{tab:roi20x20} lists CME speeds together with critical heights, mean transverse field values at different heights and ribbon flux for this ROI.  The strongest individual relationship is found for the weighted critical height, with a Pearson correlation of $r_p = 0.73$. By contrast, correlations with $\overline{B_t}$ alone are weaker, with the mean field at 150~Mm giving highest $r = 0.33$. 

\textbf{When both parameters are combined in a multiple linear regression, as defined in Eq.~\ref{eq:eq2}, the correlation improves only marginally to $r = 0.74$, indicating that critical height is the dominant predictor while $\overline{B_t}$ provides only a secondary contribution. }

Figure~\ref{fig:fig10} shows the corresponding three-dimensional plot for the best case of ROI size of $20\times20$. The blue points represent the bulk of the sample ($v < 1500$~km~s$^{-1}$), while red points mark the four outliers with CME speeds above 1500~km~s$^{-1}$. The fitted line is drawn using equation \ref{eq:eq2} and  highlights the joint dependence on weighted $h_c$ (x-axis) and $\overline{B_t}$ at 150~Mm (y-axis), with CME speed plotted on the z-axis for blue points. The regression indicates that CME speed increases with both weighted $h_c$ and $\overline{B_t}$ at 150~Mm, with the dependence on $h_c$ being substantially stronger. The Pearson correlation obtained here ($r_p$=0.73) matches the value from the PIL-only analysis in Section~\ref{sec:PIL_only}, reinforcing the consistency of the result. Critical height remains the most reliable single-parameter predictor of CME speed across events and across all tested parameters.

\begin{figure}[h]
    \centering   \includegraphics[width=\linewidth]{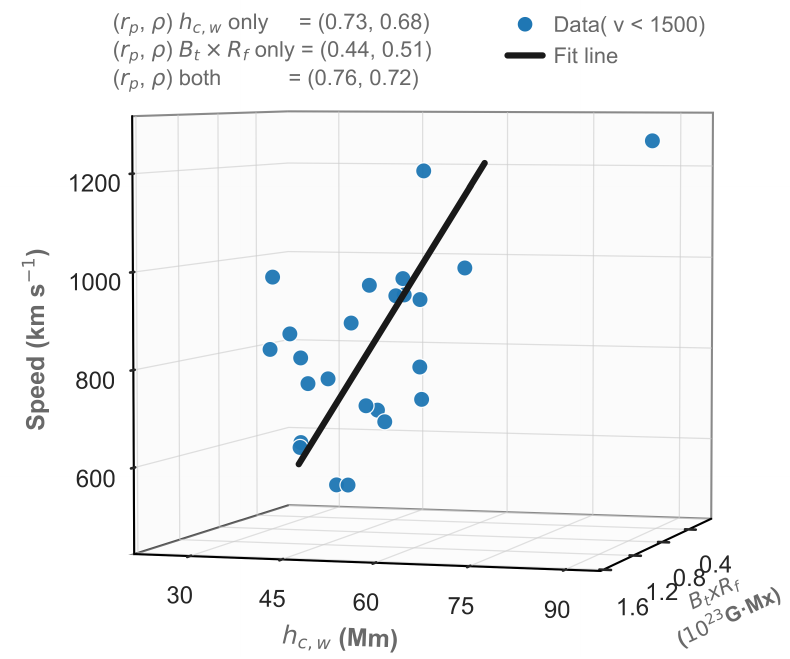}
\caption{3D plot of CME speed (km s$^{-1}$) versus weighted critical height $h_c$ and $\overline{B_t}$x$R_f$. Blue points are events with $v<1500$ km s$^{-1}$; the black line is the linear fit on $h_c$ and $\overline{B_t}$x$R_f$. Reported correlation coefficients correspond to $h_c$ alone, $\overline{B_t}$x$R_f$ alone, and $(r_p, \rho)$ for a multivariate linear model using both predictors.}
    \label{fig:fig11}
\end{figure}

For a subset of events, we also test whether including the ribbon flux from the active region closest to CME onset improves the correlation with CME speed. Ribbon flux values are taken from the catalog of \citet{kazachenko2017database}. Of our event list, 24 eruptions have associated ribbon flux measurements from the catalog, with the remaining four shown as dashed entries in Table~\ref{tab:roi20x20}. Figure~\ref{fig:fig11} shows CME speed plotted against weighted $h_c$ and the product $B_t(150~\mathrm{Mm}) \times R_f$ for these 24 events. The correlation with weighted $h_c$ alone is Pearson $r=0.73$ and Spearman $\rho=0.68$, while with $B_t \times R_f$ alone it is $r=0.44$, $\rho=0.51$. Including both parameters in a multiple regression, as defined in Eq.~\ref{eq:eq2}, improves the correlation modestly, yielding $r_p=0.76$, and $\rho=0.72$. Weighted $h_c$ again provides the stronger contribution, extending the finding from Figure~\ref{fig:fig10} and showing that while the inclusion of ribbon flux slightly improves the fit, the critical height remains the dominant predictor of CME speed.

\section{Discussion}
    \label{sec:Dis}
Critical height, derived from decay index profiles above polarity inversion lines (PILs), has previously shown a strong positive correlation with 3D CME speed, as demonstrated in Paper~I \citep{gandhi2025linking}. Building on those findings, this study expands the analysis to include both PIL-based and region-based extrapolations, and compares critical height against other physically motivated predictors. Unlike Paper~I, which used Fourier-based potential field extrapolation \citep{alissandrakis1981field} and focused solely on PIL-tracking, the present study applies Green's function-based extrapolation and introduces fixed PIL-centered regions of interest (ROIs) to evaluate multiple critical height metrics. This dual approach enables direct comparison between temporally tracked PIL segments and single-frame ROI diagnostics. Among all tested parameters, including overlying transverse field strength($B_t$) and ribbon flux($R_f$), critical height($h_c$) remains the most strongly correlated with 3D CME speed (see dark grey bars in Figure~\ref{fig:fig9}. The ROI-based weighted $h_c$ from a $20\times20$ region yields $r = 0.73$ for CMEs with speeds below $1500$kms$^{-1}$. For the PIL-based method, the correlation is $r_p = 0.71$ when all 28 events are included, but it increases to $r_p = 0.73$ when restricted to the $v < 1500$kms$^{-1}$ subset, indicating that both approaches give consistent results within this range. 

Since Paper I analyzed 37 events, all with speeds below $1500$kms$^{-1}$, and reported Pearson and Spearman correlations of $0.71 \pm 0.08$ using a similar automated PIL-tracking method with Fourier-based extrapolation \citep{alissandrakis1981field}, we restrict our comparison to the same speed range. Our correlation of $r = 0.73$ is slightly higher but well within the reported uncertainty of Paper~I. This modest increase may reflect two factors: the smaller sample size considered here, and the use of Green’s-function extrapolation, which avoids the periodic boundary assumptions of the Fourier method and may better capture local magnetic structure near the PIL. Together, these results confirm that $h_c$ is a consistent and robust single-parameter predictor of CME speed across different extrapolation techniques. They also improve upon earlier studies that used plane-of-sky CME speeds and reported weaker correlations (typically $r \lesssim 0.6$) with magnetic parameters \citep[e.g.,][]{xu2012relationship,deng2017roles}.

Compared to $h_c$, the overlying transverse magnetic field $B_t$ (green bars in Figure~\ref{fig:fig9}) shows weaker correlations across all tested heights, with the highest at 150\,Mm ($r_p = 0.33$). Combining $h_c$ with $B_t$ (purple bars in Figure~\ref{fig:fig9}) yields only marginal improvement ($r_p = 0.74$), reinforcing the finding that critical height alone explains most of the variance in CME speed. \textbf{This is consistent with the torus instability framework, in which the decay-index profile of the background field governs the loss of equilibrium and eruption onset \citep{kliem2006torus,demoulin2010criteria,zuccarello2015critical}. }

To assess the flux-accretion (FA) model \citep{welsch2018flux}, we tested the combined influence of $B_t$ and ribbon flux $R_f$. For each CME, the flare closest to eruption onset was identified, and the corresponding $R_f$ was taken from the same NOAA region at the nearest available time, based on the catalog of \citet{kazachenko2017database}. $B_t$ was computed from the magnetogram closest to CME onset. For the 24 events with available ribbon data, $B_t \times R_f$ correlates moderately with CME speed ($r_p = 0.44$), improving to $r_p = 0.76$ when combined with $h_c$ (see purple bars in Figure~\ref{fig:fig9}). This confirms that reconnection contributes to CME acceleration \citep{aulanier2009formation,cheng2020initiation}, but demonstrates that the background field structure, particularly the critical height $h_c$, remains the dominant factor controlling the overall correlation.

Among the 28 events analyzed, four with CME speeds exceeding $1500,\mathrm{km,s^{-1}}$ behave as outliers in the ROI-based analysis (Figure~\ref{fig:fig10}). Such fast CMEs are rare and usually originate from magnetically complex active regions, where multiple inversion lines may be present. The ROI method samples a broader set of pixels around the PIL, which can dilute the signal from the eruption site and increase scatter in these cases. These four events were excluded when computing the correlation with weighted $h_c$, yielding $r = 0.73$ for the remaining 24 events. In contrast, the PIL-only method, which tracks pixels directly along the inversion line, includes all 28 events and gives a correlation of $r = 0.71$ (see red and blue points in Figure~\ref{fig:fig8}). \textbf{For the same subset of 24 events, this increases to $r = 0.73$ (blue points in Figure~\ref{fig:fig8}), indicating that both methods capture consistent overall trends}. The main difference is that the four fastest CMEs lie closer to the regression line when critical heights are taken from the PIL-only method, suggesting that direct PIL tracking can better represent fast eruptions when the inversion line is well defined. \textbf{However, this subset of four fast CMEs is too small for statistically robust conclusions, and larger samples will be required to verify whether this behaviour is systematic.} ROI-based extrapolations, however, remain less sensitive to subjective pixel selection and provide a scalable framework for estimating critical heights in complex active regions, making them more suitable for automation and large-sample studies. 

\textbf{With a larger sample size, the overall scatter would be expected to decrease, particularly at the high-speed end, as a broader range of magnetic configurations and CME types become statistically represented. This would likely stabilise the correlation coefficients and narrow their uncertainties, allowing more robust assessment of secondary parameters such as $B_t$ and $R_f$. In other words, while the current trends are consistent and significant, a larger dataset would primarily improve confidence levels and refine regression estimates rather than alter the physical interpretation.}

Our approach also has several limitations. \textbf{Potential-field extrapolations neglect coronal currents and magnetic shear and therefore underestimate non-potential components of the field \citep{forbes2006cme,low1996solar}. This leads to systematically lower transverse field strengths and a compressed dynamic range of $B_t$ which likely weakens its correlation with CME speed ($r_p$=0.33 in Figure~\ref{fig:fig10}) compared to what might be obtained from non-linear force-free (NLFFF) models. While this limitation affects the absolute values of $B_t$, the relative trends across different heights remain meaningful for comparing the strength of overlying fields among events. The instability threshold is assumed to occur at a decay index threshold of 1.5, although values between 1.1 and 1.9 have been reported depending on geometry and flux-rope properties \citep{filippov2021critical,kliem2024decay}. Varying this threshold would shift the derived critical heights by several tens of megametres but would not alter their relative ordering among events, so the overall correlation trends would remain similar.} Critical height estimates from PIL-based extrapolation can be influenced by active-region complexity, Hale class, and projection effects near the limb. In practice, the only manual step in our method is the selection of initial active-region coordinates to create magnetogram cutouts, which could be replaced by standard data products such as HMI SHARPs to allow fully automated processing. Ribbon flux values are taken from single pre-eruption frames and do not capture temporal evolution, which could underestimate reconnection rates and therefore the apparent contribution of $B_t$x$R_f$
to CME acceleration.

Despite these constraints, the overall trends are clear. Critical height emerges as a physically motivated and observationally accessible proxy for CME speed, consistent across both PIL-based and ROI-based methods. The modest contribution of $B_t \times R_f$ supports the FA hypothesis in principle but highlights that reconnection-related parameters alone do not fully account for CME dynamics. A combined framework that includes both field topology and reconnection signatures may offer improved predictive capability.  \textbf{Incorporating non-linear force-free (NLFFF) models \citep[e.g.,][]{wiegelmann2021solar} would likely yield higher and more physically representative correlations, particularly for the transverse field component.}

\section{Conclusion}
\label{sec:conc}

We examine the relationship between CME speed and coronal magnetic diagnostics derived from potential-field extrapolations for 28 eruptive active regions. Critical height, obtained either from temporal tracking of PIL-specific decay index profiles or from single-frame ROI-based extrapolations, shows the strongest and most consistent correlation with 3D CME speed. The PIL-based method gives a correlation of $r_p = 0.78$, while the ROI-based weighted $h_c$ from a $20\times20$ region yields $r_p = 0.73$. Both values exceed the correlation reported in Paper~I ($r_p = 0.71$) and are substantially higher than those found in earlier studies based on plane-of-sky speeds. These results confirm that critical height is a robust diagnostic of CME dynamics.

By comparison, the mean transverse field strength $B_t$ at fixed coronal heights shows weaker correlations, with the strongest value of $r_p = 0.33$ at 150\,Mm. Combining $h_c$ and $B_t$ yields only marginal improvement, indicating that critical height captures the dominant contribution. Testing the flux-accretion model, we find that $B_t \times R_f$ correlates moderately with CME speed ($r_p = 0.44$), improving to $r_p = 0.76$ when combined with $h_c$. This suggests that reconnection contributes to acceleration but that $h_c$ provides the stronger predictive constraint.

Outliers with CME speeds above 1500\,km\,s$^{-1}$ deviate from the general trend and reduce the overall correlation, yet when excluded, both PIL- and ROI-based methods give consistent results ($r_p = 0.73$). This agreement highlights ROI analysis as a scalable alternative to PIL-tracking, reducing subjectivity in PIL selection and providing a standard reference for large-sample or operational contexts. Although PIL-specific measurements remain slightly more effective for the fastest CMEs, ROI-based diagnostics reproduce the main trends and are less sensitive to detection biases.

In summary, critical height emerges as the most reliable single predictor of CME speed in our sample, with ROI-based methods offering a practical and automatable approach when multiple or complex PILs are present. The inclusion of ribbon flux provides secondary improvement, while mean transverse field strengths alone show limited predictive value. Together, these results support the use of critical height as a key parameter for linking active-region magnetic structure to CME dynamics and demonstrate a pathway toward automated magnetic diagnostics suitable for space weather forecasting applications.

\section*{Acknowledgments}
H.G. is supported by the Science and Technology Facilities Council (STFC) through a PhD studentship at Aberystwyth University. H.M. acknowledges support from STFC grant UKRI1213: CorMag to Aberystwyth University. This work uses data from the Helioseismic and Magnetic Imager (HMI) onboard the Solar Dynamics Observatory (SDO), courtesy of NASA/SDO and the HMI science team, with access provided through the Joint Science Operations Center (JSOC) at Stanford University. Flare ribbon data are taken from the database of \citet{kazachenko2017database}. We thank Alex James, Lucie Green, and Brian Welsch for helpful discussions and valuable suggestions that improved this work.
\balance
\bibliography{references}
\bibliographystyle{aasjournal}
\end{document}